\magnification=1200
{\bf Finite Size Scaling for First Order Transitions: Potts Model}
\bigskip
P.M.C. de Oliveira, S.M. Moss de Oliveira, C.E. Cordeiro and D. Stauffer$^*$,
\bigskip
Institute of Physics, Fluminense Federal University\par
Av. Litor\^anea s/n, Boa Viagem, Niter\'oi, RJ 24210-340, Brazil
\bigskip
{\bf Abtract}

The finite-size scaling algorithm based on bulk and surface renormalization
of de Oliveira (1992) is tested on $q$-state Potts models in dimensions $D =
2$ and 3. Our Monte Carlo data clearly distinguish between first- and
second-order phase transitions. Continuous-$q$ analytic calculations
performed for small lattices show a clear tendency of the magnetic exponent
${\cal Y} = D - \beta/\nu$ to reach a plateau for increasing values of $q$,
which is consistent with the first-order transition value ${\cal Y} = D$.
Monte Carlo data confirm this trend.

\bigskip
{\bf Introduction}

A new type of renormalization or finite-size scaling algorithm was proposed
some time ago [1] and tested on a variety of models [2]. For spin 1/2 Ising
models, it is based on a majority rule for either all spins of the system or
the spins on two opposite surfaces of the finite system under consideration.
Basically, the surface criterion tests whether the two opposite surfaces have
predominantly the same magnetization or are mostly uncorrelated. In this way
one finds out whether the system is in a single-domain state or is divided
into many domains with different signs of the magnetization.

In the $q$-state Potts model [3], each spin can be in one of $q$ different
states, $q = 2$ corresponding to the usual Ising model. Accordingly, we now
check for the two opposite surfaces of the lattice, which spin state
dominates one surface and which the other one. If both surfaces are dominated
by the same state, a counter is increased by 1; if they are dominated by two
different states, this counter is decreased by $1/(q-1)$. If there is no
correlation between the two surfaces, the average value of this counter thus
stays at zero, whereas for well-correlated surfaces due to long-range
ferromagnetic order, the counter is increased by one unit each time it is
measured. At the end, we normalize the counter by the number of measurements,
denoting the result by ${\cal T}$; in the thermodynamic limit, ${\cal T}$ is
unity for well-correlated surfaces (temperature $T < T_c$) and zero for
uncorrelated ones ($T > T_c$), i.e. a step function of $T-T_c$. We start with
a random distribution of spins, wait until equilibrium has been established
at the given temperature $T > T_c$, then decrease the temperature slightly
for a new equilibrium situation, and so on until $T < T_c$ is reached.

In this form this criterion holds for second-order transitions where the
correlation length diverges. At a first-order transition, however, the
correlation length remains finite, and thus a multi-domain state is possible
also below the phase transition temperature, if all coexisting states have the
same energy. This is the case for the $q$-state Potts model in two dimensions
if $q$ is larger than 4, and in three dimensions, if $q$ is at least 3. We thus
expect in a simulation below the transition temperature, that for a
second-order transition the normalized counter is close to one for all samples,
whereas for a first-order transition it is close to zero in some samples and
close to one in others, for lattice sizes not larger than the correlation
length.

The complications inherent to the first-order transition can be avoided if,
instead of cooling down slowly from a random initial configuration, we heat
up slowly an initially ordered configuration where all spins are the same.
Then, below the transition temperature the normalized counter should stay
near unity, and should jump to zero if we heat above the transition
temperature.

We have written a C-language program which stores eight lattices in 32-bit
words, allowing $0 < q < 16$. We ran it on PC's for testing in two dimensions,
and on IBM Powerstations for runs with up to $658 \times 658$ and $106 \times
106 \times 106$ spins. Sites were updated regularly; for each Monte Carlo
updating a new direction for the spin was selected randomly and accepted with
the usual thermal probability of the Metropolis technique. As we adopted
periodic boundary conditions, the above-mentioned two opposing surfaces of an
$L \times L \times L$ cubic lattice correspond to two parallel planes $L/2$
lattice parameters apart from each other. In order to improve the statistics,
after each whole lattice sweep (WLS), we average correlations between planes
$i$ and $i+L/2$ for $i = 1, 2 \dots L/2$ along all three directions,
corresponding to a total of $3L/2$ distinct averaged values.

\bigskip
{\bf Qualitative Results}

In three dimensions, our data qualitatively confirm our two-dimensional
conclusions to be presented in the next paragraph: for $q = 2$ (Ising model),
the transition was found near $J = 0.22$ (in units of $k_BT$), and for $q=4$
near $J = 0.15$. In the Ising case, at low temperatures the normalized
counter was unity in all eight simulated lattices. The same applied for $q =
4$ only if we started with an ordered phase and low temperatures and heated
the system up. If instead we cooled down an initially random configuration,
then some lattices had the normalized counter at or near unity, and the
others near zero.

In two dimensions, Fig. 1 shows this behavior for $q = 5$ on an $L \times L$
lattice with $L = 32$: we simulated 16 samples, starting from 16 distinct
random spin configurations at high temperatures (low values of the coupling
constant $J$ in the plot). After 1000 transient and 10000 averaging WLSs for
each fixed temperature, we decrease $T$ by a small finite amount and perform
another set of $1000+10000$ WLSs. Below the critical point, we observe that
10 among these 16 samples have become ordered, and the average value of the
counter is ${\cal T} = 1$ for them. The other 6 samples remain disordered,
the averaged value of the counter being ${\cal T} \sim 0$ for them. Taking
separate averages for these two sets of samples, we get the two curves shown
in Fig. 1. They collapse onto each other near the known transition
temperature corresponding to $J_c = 0.234872$, both staying at zero for
higher temperatures (disordered phase). We have not observed any
metastability effect for both first- and second order transitions. For
instance, we have observed no trace of hysteresis by heating or cooling (in
this case choosing only ordered low temperature samples) the same system.

As explained above, the behavior observed in Fig. 1 is expected for
first-order transitions. Nevertheless, it is dangerous to adopt this
criterion in order to determine whether a given phase transition is first- or
second-order. For instance, we get the same qualitative behavior for $q = 3$
and 4, in two dimensions, for which the transition is known to be continuous,
by using the same parameters. In the latter case, this is only a transient
time- and finite size- effect, and one can get all samples ordered at low
temperatures simply by taking longer transient times and larger lattice
sizes. However, the computer power actually needed to draw such a distinction
may become prohibitively high. Thus, our criterion to determine the character
(continuous or first-order) for a given phase transition is another
quantitative one, based on the magnetic critical exponent ${\cal Y}$ to be
explained below.

\bigskip
{\bf The Method}

The surface correlation function ${\cal T}$ has already been defined. The
bulk quantity $Q$ is defined as the average of another counter, as follows.
First, one must adopt a privileged state among the $q$ possible spin
orientations (this corresponds simply to choose a particular orientation of
an external magnetic field). For a given lattice spin configuration we
determine whether the majority of spins are in this privileged state (setting
the counter equal to 1) or in any other state (setting the counter equal to
$-1/(q-1)$, instead).  We define $Q$ as the average of this counter. For $q =
2$, this corresponds to taking the average of the sign of the sum of the
Ising spins $\pm 1$, according to the original definition [1,2].

Our method is based on these two thermodynamic quantities ${\cal T}$ and $Q$
which are shown to scale as $L^0$ [1,2] at the critical temperature and in
the thermodynamic limit, where $L$ is the linear size of the system. So, they
behave like Binder's fourth order cumulant [4] or the Nightingale ratio
(correlation length over strip width) [5], or yet to the Ziff's percolation
spanning probability [6], being step functions of the temperature $T$, in the
thermodynamic limit. Concerning the phenomenological RG [5], it has been
shown [7] to be a particular case of our method if one computes ${\cal T}$ by
using lattices with the same strip geometry. However, these functions ${\cal
T}$ and $Q$ have the further advantage of also describing correctly the
critical behavior around $T = 0$ and $T \to \infty$, including the effects of
an external magnetic field $H$ as well. For instance, according to the
definition of $Q$, one can obtain exactly the low temperature scaling
behavior $H \sim \ell^D$ for ferromagnets, the first-order character
signature of the reversing field transition below the critical temperature,
where $\ell$ is the RG length scaling factor. Also, one obtains $H \sim
\ell^{D/2}$ at high temperatures, the signature of the lack of long range order
above the critical temperature. On the other hand, in zero field, the function
${\cal T}$ gives the correct scaling behavior $J \sim \ell^{D-1}$ at low
temperatures. The function $Q$ is based on a bulk measure of the majority of
spins, whereas ${\cal T}$ measures the correlation between two opposite
surfaces of the system.

The magnetic exponent ${\cal Y}$ is obtained through the finite size scaling
relation $M \sim L^{\cal Y}$, valid at the critical temperature, where $M$ is
the bulk magnetization. For the Ising model, $M$ is simply the average of the
absolute value of the sum of all spins [1,2]. For general $q$ on a finite
lattice with $N$ spins, one must first determine the state of the majority of
the spins for the current configuration, and count the number $m$ of these
majority spins. Then, the value of $M$ will be the average of
$m - (N-m)/(q-1)$.

For a first-order transition, this quantity $M$ clearly scales as $L^D$,
where $D$ is the geometrical dimension of the system. Thus, one can extend
the definition of the magnetic critical exponent ${\cal Y}$ even to
first-order transitions. On the other hand, the thermal critical exponent
$\nu$ which governs the divergence of the correlation length cannot be
defined in those cases. Fig. 2 shows the $q$-dependence of the critical
coupling $J_c$ and the exponents ${\cal Y}$ and $\nu$, known for the
two-dimensional Potts model [3]. We performed continuous-$q$ analytic
calculations, for small lattices, which showed a clear tendency of the
magnetic exponent ${\cal Y}$ to reach a plateau for increasing values of $q$,
mimicking the correct behavior of Fig. 2, but without the jump at $q = 4$.
These preliminary results motivated us to simulate larger lattices.

The critical coupling $J_c$ and the thermal exponent $\nu$ can be obtained by
plotting the function ${\cal T}$ against $L^{1/\nu} (J-J_c)$ for different
linear lattice sizes $L$, in zero field. The parameters $J_c$ and $\nu$ must
be adjusted in order to collapse all data onto the same curve. The result is
shown for the two-dimensional Ising model in Fig. 3a. One can note the good
accuracy obtained with a small computational effort. Values obtained for
larger lattices also collapse onto this same curve, but we decided to plot
only data corresponding to the two smallest lattice sizes simulated, in order
to show that one has no need of much computer power. We have chosen the
correct known value $\nu = 1$ and adjusted $J_c = 0.4407$, also coincident
with the correct known value. Any deviation on the fourth decimal clearly
gives a poorer collapse. Similar accuracy is also obtained for $q = 3$, 4, 5
and 6 in two dimensions, with the same modest computational effort. In the
cases of $q = 5$ and 6, the curves present an almost vertical jump near the
transition point, and one can extract $J_c$ without resorting to data
collapsing. Nevertheless, we obtain also a good collapse also for these cases
where the thermal exponent is not defined, adopting the value $\nu = 2/3$
corresponding to $q = 4$ (the collapse is not much sensible to this
artificial value). The case for $q = 6$ is shown in Fig. 3b.

The magnetic exponent ${\cal Y}$ is obtained by further simulations at $J_c$,
measuring the value of $M$ for different linear lattice sizes $L$. Fig. 4
shows the results for $q = 5$ in two dimensions. The straight line confirms
the expected scaling relation $M \sim L^{\cal Y}$, with ${\cal Y} = 1.997$ in
this case, corresponding to an error less than $0.2 \%$ relative to the
first-order value ${\cal Y} = D = 2$. For $q =6$, we obtained ${\cal Y} =
2.010$. Two-dimensional Potts model simulations are normally used as tests
for methods conceived to determine whether a given transition is first- or
second-order, and we are aware of this kind of work (see, for instance [8])
for at least $q =7$ and upwards where the first-order character is already
well defined. The good results we obtained for $q = 5$ and 6 give confidence
in the present method. Another very good test is the three-dimensional Potts
model for $q = 3$, known to suffer a weak first-order transition, staying
very near the borderline from second-order. From our simulations we get
${\cal Y} = 2.97$, with a similar modest computational effort to that
employed in two dimensions.

The second order transitions for $q = 2$, 3 and 4 in two dimensions seem to
need more computational effort in order to yield the same degree of accuracy
in determining ${\cal Y}$. We get ${\cal Y} = 1.86$ for $q = 2$ (statistical
errors are always in the last displayed digit), to be compared with the known
value ${\cal Y} = 1.875$. For $q = 3$, our value ${\cal Y} = 1.93$ must be
compared with ${\cal Y} = 1.867$. Even with this poorer accuracy, our data
are safe enough to guarantee that ${\cal Y} < D$ in both cases, confirming
the continuous character of these transitions. For the borderline case $q =
4$, we get ${\cal Y} = 1.997$ to be compared with ${\cal Y} = 1.875$. The $q
= 4$ case is indeed problematic not only for the present method but for all
other numerical methods, because it is known to present very strong
logarithmic  corrections to the simple scaling power laws. Moreover, in Fig.
2 one can note the gap between ${\cal Y} = 1.875$ and ${\cal Y} = 2$
occurring just at this borderline $q = 4$. Our result for $q = 4$ reproduces
very well the correct one for $q = 4 + \varepsilon$ [9].

In conclusion, the algorithm of ref. 1 seems to work also for first-order
transitions, being very accurate also in these cases.

The authors are indebted to S.L.A. de Queir\'oz for a critical reading of the
manuscript. DS thanks the Brazilian Government for a CAPES travel grant. This
work was supported also by Brazilian agencies CNPq and FINEP.

\bigskip
{\bf References}

\item{*} Present and permanent address: Institute for Theoretical Physics,
Cologne University, D-50937 K\"oln, Germany.\par

\item{[1]} P.M.C. de Oliveira, {\it Europhysics Letters} {\bf 20}, 621
(1992).\par

\item{[2]} J.M. Figueiredo Neto, S.M. Moss de Oliveira and P.M.C. de
Oliveira, {\it Physica} {\bf A206}, 463 (1994).\par

\item{} P.M.C. de Oliveira, {\it Physica} {\bf A205}, 101 (1994).\par

\item{} S.M. Moss de Oliveira, P.M.C. de Oliveira and F.C. S\'a Barreto, {\it
J. Stat. Phys.} {\bf 78}, 1619 (1995).\par

\item{[3]} R.B. Potts, {\it Proc. Camb. Phil. Soc.} {\bf 48}, 106 (1952) .\par

\item{} F.Y. Wu, {\it Rev. Mod. Phys.} {\bf 54}, 235 (1982).\par

\item{} A. Hintermann, H. Kunz and F.Y. Wu, {\it J. Stat. Phys.} {\bf 19}, 623
(1978).\par 

\item{} M.P.M. den Nijs, {\it J. Phys.} {\bf A12}, 1857 (1979).\par

\item{} B. Nienhus, E.K. Riedel and M. Schick, {\it J. Phys.} {\bf A13}, L189
(1980); R. Pearson, {\it Phys. Rev.} {\bf B22}, 2579 (1980).\par

\item{[4]} K. Binder, {\it Phys. Rev. Lett.} {\bf 47}, 693 (1981).\par

\item{[5]} M.P. Nightingale, {\it J. Appl. Phys.} {\bf 53}, 7927 (1982).\par

\item{[6]} R.M. Ziff {\it Phys. Rev. Lett.} {\bf 69}, 2670 (1992).\par

\item{[7]} J.A. Plascak and J. Kamphorst Leal da Silva, {\it preprint UFMG}
(1995).\par

\item{[8]} T. Bhattacharya, R. Lacaze and A. Morel, {\it Nucl. Phys.} {\bf
B435} [FS], 526 (1995), and references therein.

\item{[9]} R.J. Baxter, {\it J. Phys.} {\bf C6}, L445 (1973); {\it J. Stat.
Phys.} {\bf 9}, 145 (1973).\par

\bigskip
{\bf Figure Captions}

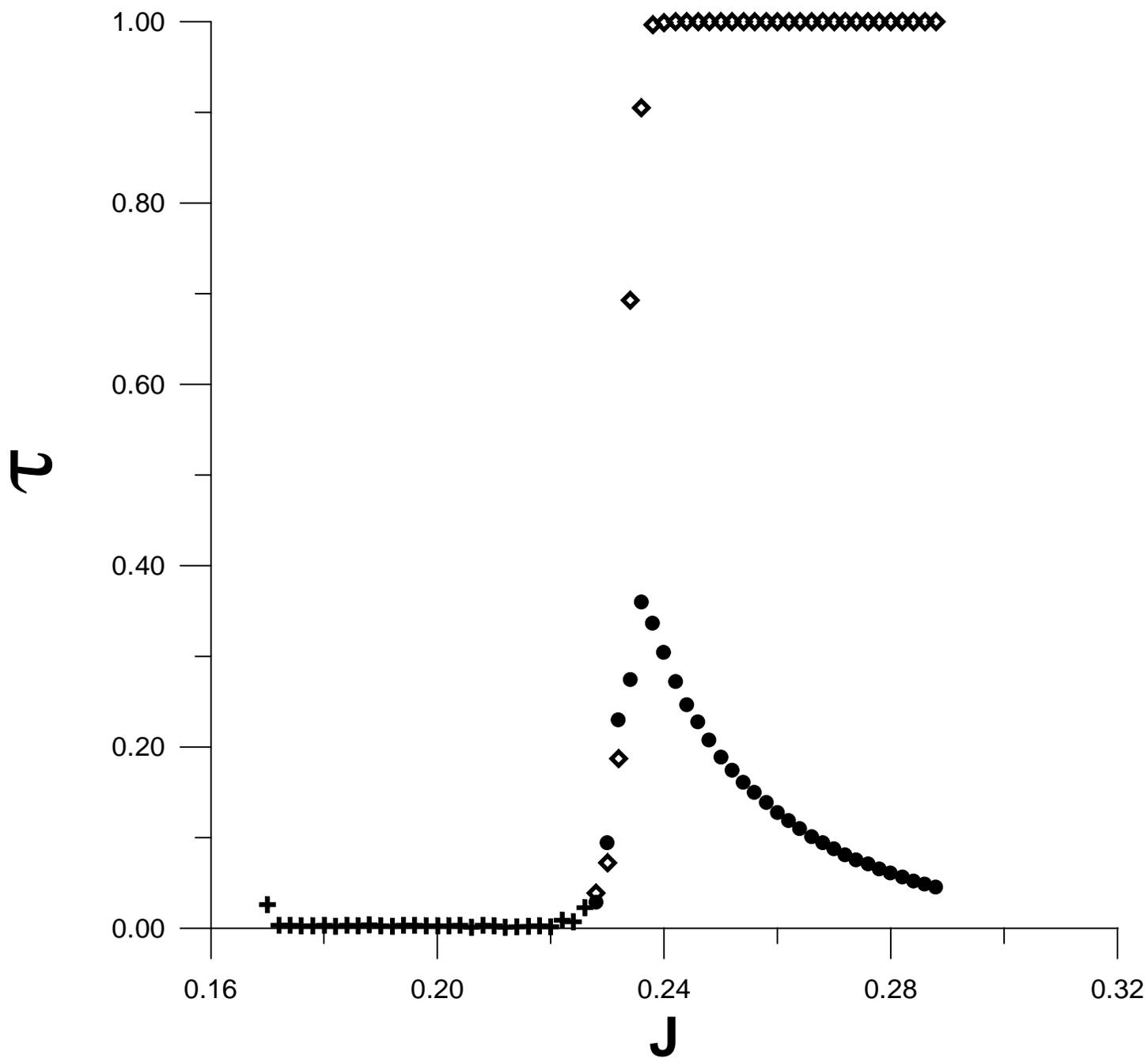
\item{Fig. 1} Surface correlation ${\cal T}$ versus coupling constant $J$ for
$32 \times 32$ lattice, and $q = 5$. Simulations start from 16 distinct
random spin configurations (samples) at a temperature well above the critical
point (at left on the $J$ axis). Diamonds correspond to the average over 10
samples which showed ${\cal T} = 1$ below the critical temperature, while
bullets correspond to the remainder 6 samples which present ${\cal T} \sim
0$. For these, the observed peak comes from the large value (although finite)
of the correlation length near the transition point. Both curves collapse
near and above the critical temperature (crosses).\par

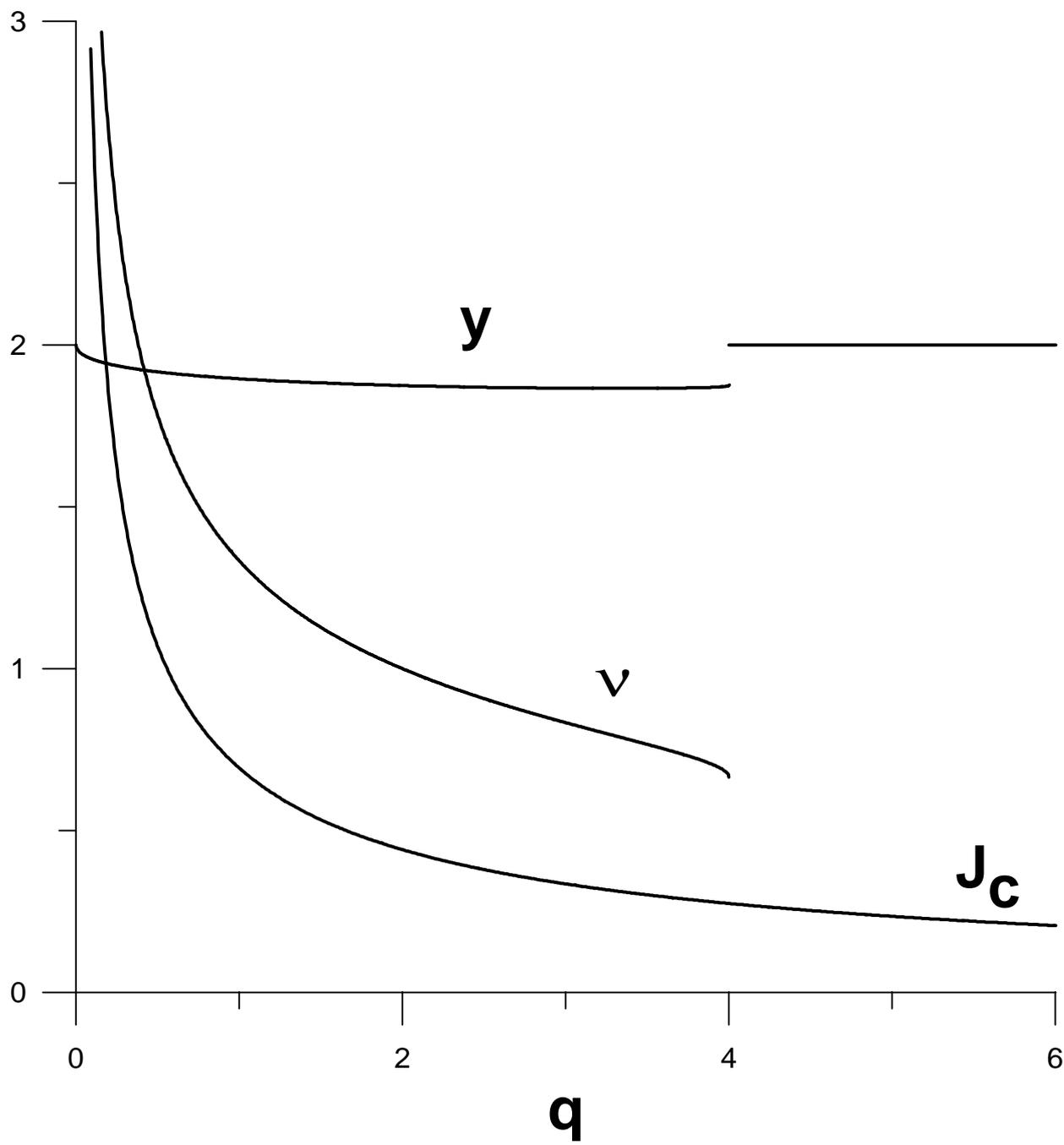
\item{Fig. 2} Exactly known values of the critical coupling $J_c$, thermal
and magnetic exponents $\nu$ and ${\cal Y}$, as functions of $q$, for the
two-dimensional Potts model. For $q > 4$, one gets ${\cal Y} = D = 2$ according
to the first-order character of the transition.\par

\item{Fig. 3} Data collapsing plots of ${\cal T}$ versus $L^{1/\nu} (J-J_c)$,
taken for $q = 2$ (a) and 6 (b) on two-dimensional lattices of linear sizes
$L = 26$ (triangles) and 34 (circles). The error bars are smaller than the
symbols.\par

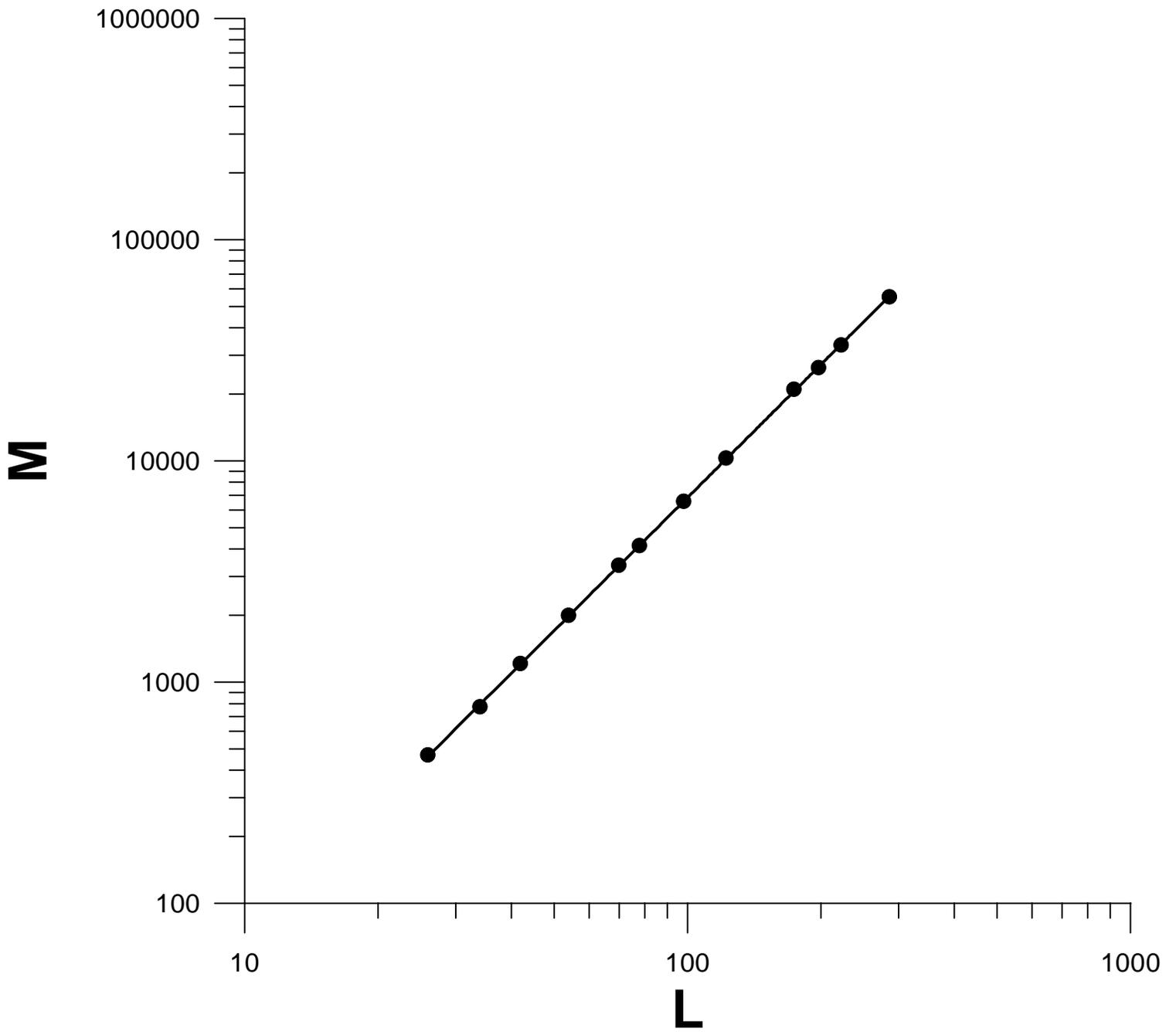
\item{Fig. 4} Plot of the bulk magnetization $M$ versus linear lattice size
$L$, measured at the critical temperature, for $q = 5$ in two dimensions. The
slope of the straight line is the magnetic exponent ${\cal Y}$.\par
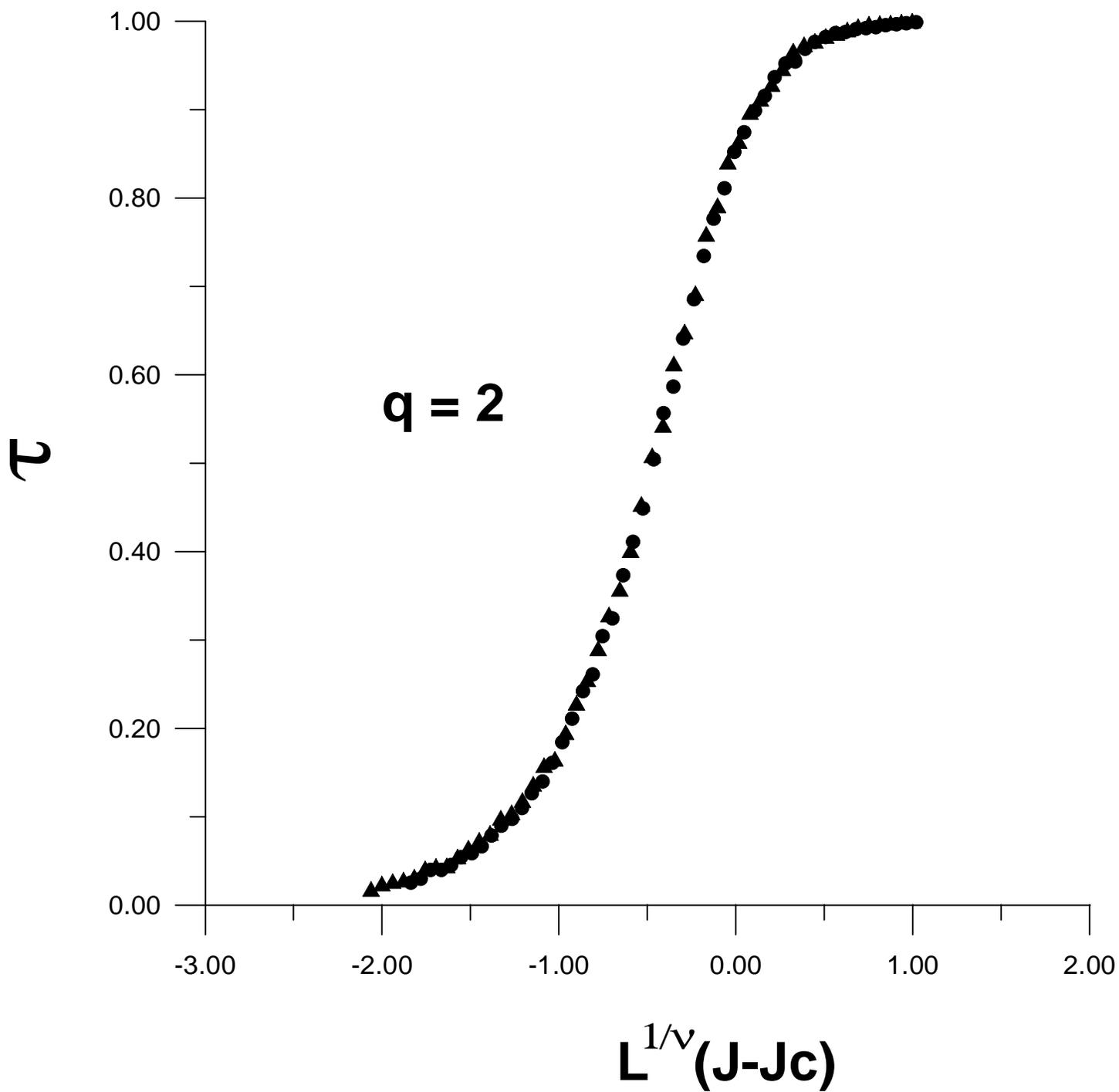
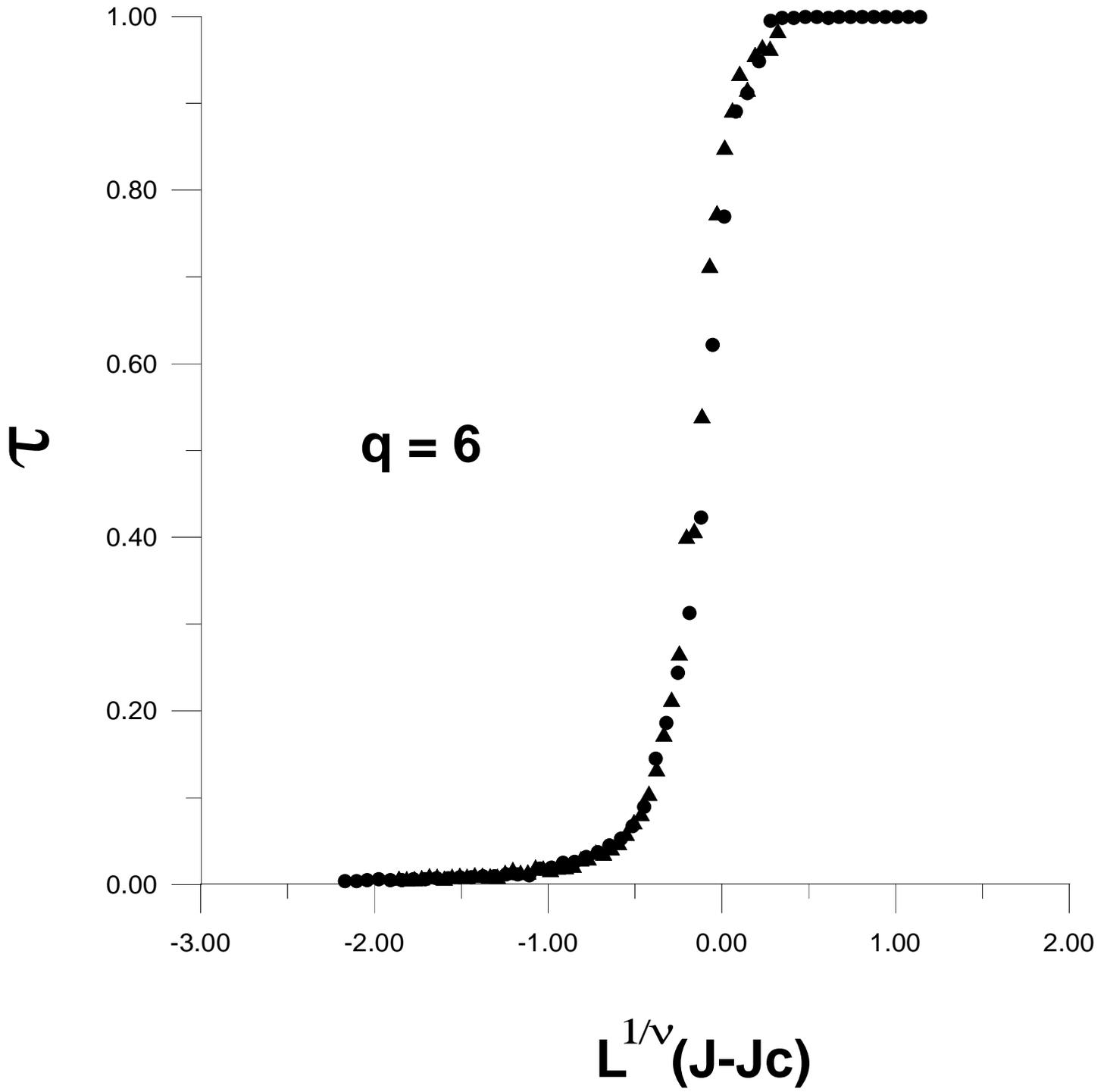
\bye